\documentclass[numberedappendix]{emulateapj}
\usepackage{apjfonts,mathptmx}

\begin{document}

\title{Spectral Methods for Time-Dependent Studies of Accretion Flows. II.\\
       Two-Dimensional Hydrodynamic Disks with Self-Gravity}

\author{Chi-kwan Chan, Dimitrios Psaltis\altaffilmark{1}, and Feryal \"Ozel}

\affil{Physics Departments, University of Arizona, 1118 E.\ 4th St., Tucson, AZ 85721}

\altaffiltext{1}{Also Astronomy Department, University of Arizona}

\begin{abstract}
  Spectral methods are well suited for solving hydrodynamic problems
  in which the self-gravity of the flow needs to be considered.
  Because Poisson's equation is linear, the numerical solution for the
  gravitational potential for each individual mode of the density can
  be pre-computed, thus reducing substantially the computational cost
  of the method.  In this second paper, we describe two different
  approaches to computing the gravitational field of a two-dimensional
  flow with pseudo-spectral methods.  For situations in which the
  density profile is independent of the third coordinate (i.e., an
  infinite cylinder), we use a standard Poisson solver in spectral
  space.  On the other hand, for situations in which the density
  profile is a delta function along the third coordinate (i.e., an
  infinitesimally thin disk), or any other function known a priori, we
  perform a direct integration of Poisson's equation using a Green's
  functions approach.  We devise a number of test problems to verify
  the implementations of these two methods.  Finally, we use our
  method to study the stability of polytropic, self-gravitating disks.
  We find that, when the polytropic index $\Gamma$ is $\le 4/3$,
  Toomre's criterion correctly describes the stability of the disk.
  However, when $\Gamma > 4/3$ and for large values of the polytropic
  constant $K$, the numerical solutions are always stable, even when
  the linear criterion predicts the contrary.  We show that, in the
  latter case, the minimum wavelength of the unstable modes is larger
  than the extent of the unstable region and hence the local linear
  analysis is inapplicable.
\end{abstract}



\section{Introduction}\label{sec:introduction}

In the standard model of accretion disks, turbulent viscosity plays an
important role in bringing material inward and transporting angular
momentum outward \citep[see][]{Frank2002}.  At the same time, viscous
dissipation converts gravitational potential energy to thermal energy
and heats up the disk, which then radiates away this energy as thermal
emission.  In most applications of accretion disks around central
objects, as in, e.g., X-ray binaries, the self-gravity of the flow is
negligible compared to that of the central object.  However, there are
disk-like systems, such as active galactic nuclei as well as
protostellar and protoplanetary disks, where the effects of
self-gravity change not only the properties of angular momentum
transport \citep{Boss1998, Balbus1999, Mejia2005}, but also the energy
balance equation \citep{Bertin1997,Bertin1999,Bertin2001}, which
affect the global structure of the disk.

Besides the angular momentum transport problem, self-gravitating disks
are also very important in studying star and planet formation.
Indeed, gravitational instabilities in protoplanetary disks have been
proposed as viable planet formation mechanisms.  Although there has
been a large amount of work done on gravitational instabilities in
these system \citep[see, for example,][and references
therein]{Pickett2003,Mejia2005}, it still remains an open question
whether the fragmentation by gravitational instabilities can produce
bound planetary objects, or if reservoirs of small solid cores are
needed for the accretion mechanism to lead to rapid planet formation.

In the first paper in this series (Chan, Psaltis, \& \"Ozel 2005), we
presented a pseudo-spectral method for solving the equations that
describe the evolution of two dimensional, viscous hydrodynamic flows.
There, we addressed issues related to the implementation of boundary
conditions, spectral filtering, and time-stepping in spectral methods,
and verified our algorithm using a suite of hydrodynamic test
problems.  In this second paper of the series, we present our
implementation of a Poisson solver that allows us to take into account
the effects of self-gravity of the flow.

Spectral methods are particularly suitable for incorporating the
effects of self-gravity.  Because Poisson's equation is linear, we can
pre-compute the numerical solution for the gravitational potential for
each individual mode of the density, thus reducing substantially the
computational cost of the method.  In fact, Fourier methods have been
used extensively in analytical studies of gravitational potentials in
systems with periodic boundary conditions \citep{Binney1987}.  On the
other hand, in numerical studies of self-gravitating disks, the
strengths of spectral methods have only been partially incorporated.

Hybrid hydrodynamic algorithms with self-gravity have been developed
in such a way that modified spectral methods are used only for solving
Poissson's equation for the gravitational field, whereas the
hydrodynamic parts are still treated with finite difference schemes.
For example, \citet{Boss1992} describe a spherical harmonic
decomposition method and a second-order scheme in the radial direction
to solve Poisson's equation, whereas \citet{Myhill1993} use a modified
Fourier method to find the gravitational potential of an isolated
distribution of sources.  Both algorithms employ explicit second-order
finite difference methods to advance the hydrodynamic equations.
\citet{Pickett1998,Pickett2000} describe an implementation of an
algorithm that uses Fourier decomposition in the azimuthal direction
of cylindrical coordinate to solve Poisson's equation together with a
von Neumann \& Richtmeyer AV scheme for the hydrodynamics.  Some other
examples can be found in Grandclement et al. (2001), Broderick \&
Rathore (2004), and Dimmelmeier et al. (2005).  The main advantage of
using hybrid methods is that one can employ currently available
hydrodynamic algorithms based on finite difference schemes.  However,
a hybrid method does not exploit the high order of the spectral
algorithm, because the hydrodynamic difference schemes typically have
a much lower order compared to that of the Poisson solver.  Contrary
to these efforts, our algorithm uses a spectral decomposition method
for solving both the hydrodynamic and Poisson's equation, providing a
consistent treatment of the whole problem.  To our knowledge, this is
the first time that spectral methods have been used in studying
astrophysical disks with self gravity.

By construction, there is an ambiguity in the definition of the
gravitational field in two-dimensional problems.  We can assume either
that the density profile is independent of the third coordinate (i.e.,
an infinite cylinder) or that it is a delta (or any other
predetermined) function along the third coordinate (i.e., an
infinitesimally thin disk); the resulting gravitational field on the
two-dimensional domain of solution will be different in the two
cases.  For example, the gravitational potential of a ``point source''
on the two-dimensional domain of solution will be proportional to
$\log(r)$ in the first case and to $1/r$ in the second, where $r$ is
the distance from the source.  In order to consider both geometries,
here we describe two different approaches to computing the
gravitational field of a two-dimensional hydrodynamic flow with
pseudo-spectral methods.  When the flow has the geometry of an infinite
cylinder, we use a standard two-dimensional pseudo-spectral Poisson
solver, which has been proven to be numerically stable and accurate.
When the flow has the geometry of an infinitesimally thin disk, we
perform a direct integration of the Green's function for the
gravitational potential, following the work of Cohl \& Tohline (1999).

In the following section, we present our assumptions and equations.
In \S\ref{sec:numerical_methods}, we discuss the details of our
numerical methods, include both the standard Poisson solver and the
Green's function integrator.  Next, we present a series of tests in
\S\ref{sec:tests}, to verify our algorithm.  Finally, we apply our
method to a numerical study of Toomre's stability criterion of
self-gravitating disks in \S5.

\section{Equations and Assumptions} \label{sec:equations_assumptions}

We consider two-dimensional, viscous, compressible flows.  In this
second paper, we include self-gravity and continue to neglect the
magnetic fields of the flows.  The hydrodynamic equations, therefore,
contain the continuity equation
\begin{equation}
  \frac{\partial\Sigma}{\partial t} + \nabla\cdot(\Sigma\mathbf{v}) = 0, \label{eq:continuity}
\end{equation}
the Navier-Stokes equation
\begin{equation}
  \Sigma\frac{\partial\mathbf{v}}{\partial t} + \Sigma(\mathbf{v}\cdot\nabla)\mathbf{v}
  = -\nabla P + \nabla\mathbf\tau + \Sigma\;\mathbf{g}, \label{eq:navier_stokes}
\end{equation}
and the energy equation
\begin{equation}
  \frac{\partial E}{\partial t} + \nabla\cdot(E\mathbf{v}) = - P\,\nabla\cdot\mathbf{v} + \Phi - \nabla\cdot\mathbf{q}
  - \nabla\cdot\mathbf{F} - 2F_z.  \label{eq:energy}
\end{equation}
We denote by $\Sigma$ the height-integrated density, by $\mathbf{v}$
the fluid velocity, and by $E$ the thermal energy.  In the
Navier-Stokes equation, $P$ is the height-integrated pressure,
$\mathbf\tau$ is the viscosity tensor, and $\mathbf{g}$ is the
gravitational acceleration.  We use $\Phi$ to denote the viscous
dissipation rate, $\mathbf{q}$ to denote the heat flux vector, and
$\mathbf{F}$ to denote the radiation flux on the $r$-$\phi$ plane.
The last term in the heat equation, $2F_z$, takes into account the
radiation losses in the vertical direction.  The analytical forms of
the various physical quantities in
equations~(\ref{eq:continuity})--(\ref{eq:energy}) are given in
\citet{Chan2005}, except for the gravitational acceleration
$\mathbf{g}$, for which we need to introduce a new equation.

In Newtonian gravity, the gravitational field $\mathbf g$ is
conservative, so we can define the gravitational potential $\Psi$ by
\begin{equation}
  \mathbf{g} \equiv -\nabla\Psi.
\end{equation}
The gravitational potential associated with the (three-dimensional)
mass density, $\rho$, is given by the volume integral
\begin{equation}
  \Psi(t,\mathbf{x}) = -G\int\frac{\rho(t,\mathbf{x'})}{|\mathbf{x} - \mathbf{x'}|}d^3x' \label{eq:integrator_3d_total}
\end{equation}
over all space, where $G$ is the gravitational constant.  Rewriting
equation~(\ref{eq:integrator_3d_total}) in differential form, we
obtain Poisson's equation
\begin{equation}
  \nabla^2\Psi = 4\pi G\rho,
\end{equation}
with $\Psi$ satisfying the boundary condition $\Psi(t,\infty) = 0$ at
all times.  When simulating astrophysical flows, the computational
domain $\mathcal{D}^{(3)}$ is usually finite.  Based on its linearity,
we can decompose the Poisson equation into two parts
\begin{equation}
  \nabla^2\Psi_\mathrm{int} = 4\pi G\rho_\mathrm{int}, \label{eq:possion_int}
\end{equation}
and
\begin{equation}
  \nabla^2\Psi_\mathrm{ext} = 4\pi G\rho_\mathrm{ext},
\end{equation}
where $\rho_\mathrm{int}$ denotes the mass density within the
computational domain, which in our case is the flow density, and
$\rho_\mathrm{ext}$ refers to external sources such as a central
object and/or a companion star.  The gravitational field is then given
by
\begin{equation}
  \mathbf{g} = \mathbf{g}_\mathrm{int} + \mathbf{g}_\mathrm{ext} = -\nabla(\Psi_\mathrm{int} + \Psi_\mathrm{ext}).
  \label{eq:total_g}
\end{equation}

In the astrophysical context of interest here, $\Psi_\mathrm{ext}$ is
usually generated by a set of spherical objects.  Hence the external
gravitational potential is given by
\begin{equation}
  \Psi_\mathrm{ext}(t,\mathbf x) = \sum_i \frac{GM_i}{|\mathbf x - \mathbf x_i(t)|},
\end{equation}
where $M_i$ and $\mathbf x_i(t)$ are the mass and positions of the
corresponding objects.  Regarding the self-gravity of the flow, solving
equation~(\ref{eq:possion_int}) within $\mathcal{D}^{(3)}$ is
equivalent to computing the integral
\begin{equation}
  \Psi_\mathrm{int}(t,\mathbf{x}) = -G\int_{\mathcal{D}^{(3)}}
  \frac{\rho_\mathrm{int}(t,\mathbf{x'})}{|\mathbf{x} - \mathbf{x'}|}d^3x'.  \label{eq:integrator_3d}
\end{equation}
Once $\Psi_\mathrm{ext}$ and $\Psi_\mathrm{int}$ are obtained, we can
then use equation~(\ref{eq:total_g}) to obtain the total gravitational
field and subsequently use it in both the Navier-Stokes equation and
in integrating the trajectories $\mathbf x_i(t)$ for the external
objects.  Although for our test problems we assume that the central
object does not move and solve the hydrodynamic equations in a fixed
reference frame, it is trivial to generalize our algorithm to co-moving
coordinates.

There are two different approaches to reducing the above
three-dimensional formalism to two dimensions, depending on the
physical problem under study.  The first one assumes that the density
is independent of the vertical coordinate $z$, i.e.,
$\rho_\mathrm{int}(t,r,\phi,z) \equiv \rho_\mathrm{ind}(t,r,\phi)$.  In
this case, we define $\psi(t,r,\phi) \equiv
\Psi_\mathrm{int}(t,r,\phi,z)$ and we are left with a two-dimensional
problem.  We obtain the gravitational potential by solving the
two-dimensional Poisson's equation \begin{equation}
  \nabla^2\psi \equiv \left(\frac{\partial^2}{\partial r^2} + \frac{\partial}{r\partial r}
  + \frac{\partial^2}{r^2\partial\phi^2}\right)\psi = 4\pi G\rho_\mathrm{ind}
\end{equation}
\citep[This approach is used in ZEUS-2D, see][for details.]{Stone1992}

The second approach assumes that the vertical structure of the density
is described by some function $Z(r,z)$ that is independent of $t$ and
$\phi$, i.e., $\rho_\mathrm{int}(t,r,\phi,z) \equiv
\Sigma(t,r,\phi)Z(r,z)$.  We are interested in the gravitational
potential on the $z=0$ plane, so that we solve for $\psi(t,r,\phi) =
\Psi_\mathrm{int}(t,r,\phi, z=0)$.  This is a
``pseudo-two-dimensional'' problem, where the potential is not
described by a two-dimensional Poisson equation.  In order to compute
the potential properly, the easiest method, in this case, is to
integrate directly the equation
\begin{equation}
  \psi(t,r,\phi) = \int_{\mathcal{D}^{(2)}}\mathcal{G}(r,\phi;r',\phi')\Sigma(t,r',\phi')r'dr'd\phi',
  \label{eq:integrator_2d}
\end{equation}
where $\mathcal{D}^{(2)}$ denotes the two-dimensional computational
domain and
\begin{equation}
  \mathcal{G}(r,\phi;r',\phi') \equiv -G\int_{-\infty}^{\infty}
  \frac{Z(r',z')dz'}{\sqrt{r^2 + r'^2 - 2rr'\cos(\phi-\phi') + z'^2}} \label{eq:green_2d}
\end{equation}
is the ``modified Green's function'' for our problem.


\section{Numerical Methods}\label{sec:numerical_methods}

In this section, we describe the numerical methods we use to solve the
two classes of self-gravity problems.  As in the first paper
\citep{Chan2005}, we use pseudo-spectral methods, in which we expand
all functions in series by choosing the truncated functions to agree
with the approximated functions exactly at the grid points.

We use the modified Chebyshev collocation method along the radial
direction and the Fourier collocation method along the azimuthal
direction.  Let $N+1$ and $M$ be the number of collocation points in
the radial and azimuthal direction, respectively.  For every function
$f(r,\phi)$, we use the subscript $m$ to indicate the Fourier
coefficients
\begin{equation}
  f(r,\phi) = \sum_{m=-M/2}^{M/2-1} \hat f_m(r)e^{im\phi}
\end{equation}
and the subscript $n$ to indicate the Chebyshev-Fourier coefficients
\begin{equation}
  f(r,\phi) = \sum_{n=0}^N \sum_{m=-M/2}^{M/2-1} \check f_{nm}T_n(\bar r)e^{im\phi},
\end{equation}
where $\bar r \in [-1,1]$ denotes the standardized coordinate \citep[see][for notation]{Chan2005}.

We implement a standard two-dimensional pseudo-spectral Poisson
solver, which we describe briefly in \S\ref{sec:poisson_solver}
\citep[a good introduction is available in][]{Trefethen2000}.  For the
direct integrator, we describe in \S\ref{sec:direct_integrator} how to
generalize the compact cylindrical Green's function expansion
\citep{Cohl1999} with pseudo-spectral methods.  Both methods use
pre-computed matrices to speed up the algorithms.  The run-time
computational cost for both methods are of order $\mathcal{O}(N^2M +
NM\log_2M)$.

\subsection{Two-Dimensional Pseudo-Spectral Poisson Solver} \label{sec:poisson_solver}

We describe here a two-dimensional, pseudo-spectral Poisson solver.
We use the fact that the basis polynomials in the $\phi$ direction,
$e^{im\phi}$, are also eigenfunctions of the operator
$\partial^2/\partial\phi^2$ in order to split the Poisson's equation
to
\begin{equation}
  \Delta_m\hat\psi_m \equiv \left(\frac{\partial^2}{\partial r^2} + \frac{\partial}{r\partial r}
  - \frac{m^2}{r^2}\right)\hat\psi_m = \hat{f}_m \label{eq:possion_m}
\end{equation}
for $m = 0,1,\dots,M/2$, where we have set $f \equiv 4\pi
G\rho_\mathrm{ind}$.  Note that, in discrete Fourier transforms, $\hat
f_0$ and $\hat f_{M/2}$ are real but the other coefficients are
complex, so there are, in total, $M$ independent equations.  We use
$\Delta_m$ to denote the differential operator for each $m$.  In the
case of the $r-$direction, the Chebyshev polynomials are not
eigenfunctions of $\Delta_m$.  Nevertheless, because $\Delta_m$ is
time independent and linear, we can write $\Delta_m$ in matrix
representation and pre-compute its inverse $\Delta_m^{-1}$ with an
$\mathcal{O}(N^2)$ solver for each $m$.

Let $\bar r$ be the standardized coordinate of Chebyshev polynomials
and $r = g(\bar r)$ be the mapped (physical) coordinate
\citep[see][for more details]{Chan2005}.  It is well known that the
spectral Chebyshev derivative in the standardized coordinate can be
written in matrix representation as
\begin{equation}
  \bar D_{ij} = \left(\begin{array}{c|ccc|c}
  \frac{2N^2+1}{6} & \cdots & 2\frac{(-1)^j}{1-x_j} & \cdots & \frac{1}{2}(-1)N \\\hline
  \vdots & \ddots & & \frac{(-1)^{i+j}}{x_i-x_j} & \vdots \\
  -\frac{1}{2}\frac{(-1)^i}{1-x_i} & & \frac{-x_j}{2(1-x_j^2)} & & -\frac{1}{2}\frac{(-1)^{N+i}}{1+x_i} \\
  \vdots & \frac{(-1)^{i+j}}{x_i-x_j} & & \ddots & \vdots \\\hline
  -\frac{1}{2}(-1)N & \cdots & -2\frac{(-1)^{N+j}}{1+x_j} & \cdots & -\frac{2N^2+1}{6}
  \end{array}\right),
\end{equation}
which is a $(N+1)\times(N+1)$ matrix \citep[discussion of computing
this matrix accurately can be found in][]{Baltensperger2003}.  The
derivative in the mapped coordinate can be obtained by the chain rule
\begin{equation}
  \frac{\partial}{\partial r} = \frac{d\bar r}{dr}\frac{\partial}{\partial\bar r}
  = \frac{1}{dg/d\bar r}\frac{\partial}{\partial r},
\end{equation}
so that the derivative matrix in the mapped coordinate is given by the
product
\begin{equation}
  D_{ij} \equiv \frac{1}{(dg/d\bar r)|_{\bar r_i}}\bar D_{ij}.
\end{equation}
With this we are able to write
\begin{equation}
  \Delta_{m,ij}
  = \sum_k D_{ik}D_{kj} + \sum_k \frac{\delta_{ik}}{x_i}D_{kj} - \frac{\delta_{ij}}{x_i^2}m^2.
\end{equation}
The matrix $\Delta_{m,ij}$ is, of course, singular because a unique
solution to $\psi_m$ does not exist until the boundary conditions are
given.

An important case is the vanishing boundary condition
$\hat\psi_m(r_{\min}) = \hat\psi_m(r_{\max}) = 0$.  Let
$\tilde\Delta_{m,ij}$ be the $(N-1)\times(N-1)$ sub-matrix of
$\Delta_{m,ij}$ generated by removing the first and last rows and
columns.  Imposing the vanishing boundary condition to
$\Delta_{m,ij}^{-1}$ is then equivalent to finding the inverse of the
$\tilde\Delta_{m,ij}$ and filling back the first and last rows and
columns with zeros, i.e.,
\begin{equation}
  \Delta^{-1(0)}_{m,ij} = \left(\begin{array}{c|ccc|c}
  0 & 0 & \cdots & 0 & 0 \\ \hline
  0 & & & & 0 \\
  \vdots & & \tilde\Delta_{m,ij}^{-1}& & \vdots \\
  0 & & & & 0 \\ \hline
  0 & 0 & \cdots & 0 & 0
  \end{array}\right).
\end{equation}
We can then calculate
\begin{equation}
  \hat\psi^{(0)}_m(r_i) = \sum_{j=0}^N \Delta_{m,ij}^{-1(0)}\hat f_m(r_j), \label{eq:matrix_product}
\end{equation}
where the superscript $(0)$ indicates that the solutions satisfy the
homogeneous boundary conditions for all $m$.  Taking the inverse
Fourier transform along the $\phi$-direction, we then obtain the
solution $\psi^{(0)}(r,\phi)$ which satisfies the vanishing boundary
conditions.

In order to apply more generic boundary conditions, we first look for
solutions $\hat\psi^{(1)}_m(r)$ that satisfy the homogeneous
differential equation
\begin{equation}
  \Delta_m \hat\psi^{(1)}_m = 0 \label{eq:possion_homogenous}
\end{equation}
with the proper boundary conditions.  Then it is clear that the sum
$\hat\psi^{(0)}_m + \hat\psi^{(1)}_m$ satisfies
\begin{equation}
  \Delta_m\left(\hat\psi^{(0)}_m + \hat\psi^{(1)}_m\right) = \hat f_m
\end{equation}
with the same boundary conditions.  The solutions to
equation~(\ref{eq:possion_homogenous}) are
\begin{equation}
  \hat\psi^{(1)}_m(r) = \left\{\begin{array}{lll}
    C_m\ln r  + D_m        & , & m = 0 \\
    C_m r^{m} + D_m r^{-m} & , & m = 1,2,\dots,M/2
  \end{array}\right.  \label{eq:boundary}
\end{equation}
where $C_m$ and $D_m$ are complex constants, which are fixed by the
boundary conditions.  In general, these boundary conditions depend on
azimuth, i.e.,
\begin{eqnarray}
  \psi(r_{\min},\phi) & = & \beta_\mathrm{in} (\phi), \\
  \psi(r_{\max},\phi) & = & \beta_\mathrm{out}(\phi).
\end{eqnarray}
We first take the Fourier transform of $\beta_\mathrm{in}(\phi)$ and
$\beta_\mathrm{out}(\phi)$, we solve for $C_m$ and $D_m$ using
\begin{eqnarray}
  \hat\beta_{\mathrm{in},m} & = & \left\{\begin{array}{lll}
    C_m\ln r_{\min} + D_m              & , & m = 0 \\
    C_m r_{\min}^m + D_m r_{\min}^{-m} & , & m = 1,2,\dots,M/2
  \end{array}\right.  \label{eq:boundary_in}\\
  \hat\beta_{\mathrm{out},m} & = & \left\{\begin{array}{lll}
    C_m\ln r_{\min} + D_m              & , & m = 0 \\
    C_m r_{\min}^m + D_m r_{\min}^{-m} & , & m = 1,2,\dots,M/2,
  \end{array}\right.  \label{eq:boundary_out}
\end{eqnarray}
we add these terms to $\hat\psi^{(0)}_{m}$, and finally take the
inverse Fourier transform.

We summarize the two-dimensional pseudo-spectral Poisson solver using
the following steps:
\begin{enumerate}
  \item We take the Fourier transform of each physical quantity $f$
        along the $\phi$-direction and obtain $\hat f_m$ by a fast
        Fourier transform, which is of order $\mathcal{O}(NM\log_2
        M)$.
  \item We then compute the matrix product~(\ref{eq:matrix_product})
        for each $m$, which is of order $\mathcal{O}(N^2M)$.
  \item We compute $\hat\beta_m$ by a fast Fourier transform, which is
        of order $\mathcal{O}(M\log_2 M)$.
  \item We then solve for $C_m$ and $D_m$ using
        equations~(\ref{eq:boundary_in}) and (\ref{eq:boundary_out}),
        which is of order $\mathcal{O}(M)$.
  \item We impose the boundary conditions using
        equation~(\ref{eq:boundary}), for each $m$, and obtain
        $\hat\psi_m(r_k)$, which is of order $\mathcal{O}(NM)$.
  \item Finally, we take the inverse transform of $\hat\psi_m$ to
        obtain the potential $\psi$, which is of order
        $\mathcal{O}(NM\log_2 M)$.
\end{enumerate}
Therefore, the overall computational cost is $\mathcal{O}(N^2M +
NM\log_2M)$.

\subsection{Two-Dimensional Gravity Integrator} \label{sec:direct_integrator}

\begin{figure*}
  \plotone{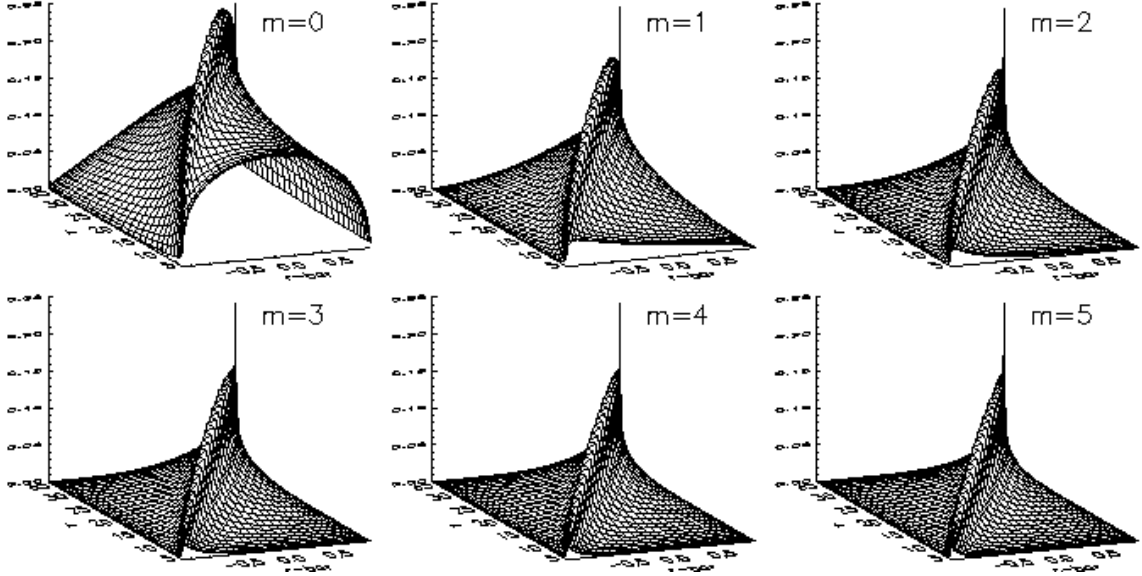}
  \caption{Plots of the quantity $\mathcal{I}_m(\bar r';r)$ for the
           first six values of $m$, when $N = 33$, $M = 64$, and
           $Z(r,z) = \delta(z)$.} \label{fig:Iknm}
\end{figure*}

As described in \S\ref{sec:equations_assumptions}, we assume in this
case that the vertical structure of the density is time-independent
and has cylindrical symmetry.  Hence we can write
\begin{equation}
  \rho_\mathrm{int}(t,r,\phi,z) = \Sigma(t,r,\phi)Z(r,z), \label{eq:restriction}
\end{equation}
where we normalize $Z(r,z)$ along the $z$ direction to unity, i.e.,
\begin{equation}
  \int_{-\infty}^\infty Z(r,z) dz = 1,
\end{equation}
so that $\Sigma$ has the physical meaning of a height-integrated
density.

We present here a new numerical method to compute the
integral~(\ref{eq:integrator_2d}) efficiently and accurately.  Our
method is motivated by \citet{Cohl1999}, who showed that the
three-dimensional Green's function in cylindrical coordinates can be
expanded in the compact form
\begin{equation}
  \frac{1}{|\mathbf{x} - \mathbf{x'}|}
  \equiv \frac{1}{\pi\sqrt{rr'}}\sum_{m=-\infty}^{\infty}e^{im(\phi-\phi')}Q_{m-1/2}(\chi), \label{eq:cohl_expansion}
\end{equation}
where $\chi\equiv[r^2 + r'^2 + (z-z')^2]/2rr'$ and $Q_{m-1/2}(\chi)$
are the half-integer degree Legendre functions.  Note that the $\phi$
and $\phi'$ dependence appears only in the exponential
$e^{im(\phi-\phi')}$.  This can be done because the Green's function
itself can be written as a function of $\Delta\phi = \phi-\phi'$ so
that $Q_{m-1/2}(\chi)/\pi\sqrt{rr'}$ is simply the Fourier transform
of $|\mathbf{x} - \mathbf{x'}|^{-1}$ with respect to $\Delta\phi$.
This property is still true for the modified Green's function because
$Z$ is $\phi$-independent:
\begin{equation}
  \mathcal{G}(r,\phi\;;r',\phi') = \mathcal{G}(r,r',\Delta\phi).
\end{equation}
Expanding it in Fourier series, i.e.,
\begin{equation}
  \mathcal{G}(r,\phi;r',\phi') = \sum_{m=-\infty}^\infty\hat\mathcal{G}_m(r,r')e^{im(\phi-\phi')},
\end{equation}
and substituting it into equation~(\ref{eq:integrator_2d}), we obtain
\begin{equation}
  \psi(r,\phi) = \int_{\mathcal{D}^{(2)}}\sum_{m=-\infty}^{\infty}\hat\mathcal{G}_m(r,r')e^{im(\phi-\phi')}
  \Sigma(r',\phi')r'dr'd\phi'.  \label{eq:integrator_2d_fourier}
\end{equation}
Note that the integration over $\phi'$,
\begin{equation}
  \int_0^{2\pi}d\phi'\Sigma(r',\phi')e^{-im\phi'} = 2\pi\Sigma_m(r'),
\end{equation}
is the Fourier transform of $\Sigma(r',\phi')$ in the azimuthal
direction.  Therefore, we can rewrite
equation~(\ref{eq:integrator_2d_fourier}) as
\begin{equation}
  \psi(r,\phi) = \sum_{m=-\infty}^{\infty}\int_{r_{\min}}^{r_{\max}}\hat\mathcal{G}_m(r,r')
  2\pi\hat\Sigma_m(r')e^{im\phi}r'dr'.
\end{equation}
Taking the Fourier transform of the whole equation again with respect
to $\phi$, we obtain a one-dimensional integral for each value of $m$
\begin{equation}
  \hat\psi_m(r) = \int_{r_{\min}}^{r_{\max}}\hat\mathcal{G}_m(r,r')2\pi\hat\Sigma_m(r')r'dr'.
  \label{eq:integrator_fourier_transformed}
\end{equation}

We now expand $\hat\Sigma_m(r')$ in Chebyshev polynomials, i.e.,
\begin{equation}
  \hat\Sigma_m(r') = \sum_{n=0}^\infty\check\Sigma_{nm}T_n(\bar r').
\end{equation}
Recalling that $\bar r'$ is the standardized coordinate, we let $r' =
g(\bar r')$ be some coordinate mapping.
Equation~(\ref{eq:integrator_fourier_transformed}) can then be
rewritten as
\begin{equation}
  \hat\psi_m(r) = \sum_{n=0}^\infty\check\Sigma_{nm}\int_{r_{\min}}^{r_{\max}}
  \hat\mathcal{G}_m(r,r')2\pi T_n(\bar r')r'dr'.
\end{equation}
Because the orthogonality condition of Chebyshev polynomials has a
weight function $(1-\bar r'^2)^{-1/2}$, i.e.,
\begin{equation}
  \langle T_l, T_n \rangle
  = \int_{-1}^{1}\frac{T_l(\bar r')T_n(\bar r')}{\sqrt{1 - \bar r'^2}}d\bar r'
  = \frac{\pi}{c_n}\;\delta_{ln},
\end{equation}
where $c_0 = 2$ and $c_n = 1$ for $n = 1, 2, \dots$, it is not
difficult to see that, by defining
\begin{equation}
  \hat\mathcal{I}_m(\bar r';r) = \hat\mathcal{G}_m\left[r,g(\bar r')\right]
  \pi\sqrt{1-\bar r'^2}\;\frac{dg^2}{d\bar r'},
\end{equation}
we can write
\begin{eqnarray}
  \hat\psi_m(r) & = & \sum_{n=0}^{\infty}\check\Sigma_{nm}
  \int_{-1}^{1}\frac{\hat\mathcal{I}_m(\bar r';r)T_n(\bar r')}{\sqrt{1 - \bar r'^2}}d\bar r' \nonumber\\
  & = & \sum_{n=0}^{\infty}\check\Sigma_{nm}\langle\hat\mathcal{I}_m, T_n \rangle \nonumber\\
  & = & \sum_{n=0}^{\infty}\frac{\check\Sigma_{nm}\check\mathcal{I}_{nm}(r)}{c_n},
  \label{eq:integrator_exact}
\end{eqnarray}
where $\check\mathcal{I}_{nm}(r)$ is the $n$-th Chebyshev coefficient
of $\hat\mathcal{I}_m(\bar r';r)$ with respect to $\bar r'$.

Up to this point, although the function $\Sigma(r',\phi')$ has been
written in Chebyshev-Fourier series, we have assumed that the series
is infinite and hence the formalism is exact.  In order to perform it
numerically, we have to discretize
equation~(\ref{eq:integrator_exact}).

For the azimuthal direction, which is periodic, we simply replace the
continuous Fourier transform by a discrete Fourier transform and use
the proper normalization, which does not change
equation~(\ref{eq:integrator_exact}) but only limits the index $m$ to
the range $-M/2$ to $M/2-1$.  For the $r$-direction, we truncate the
infinite sum at a finite number of terms, $N$.  Therefore,
equation~(\ref{eq:integrator_exact}) becomes
\begin{equation}
  \hat\psi_m(r_k) = \sum_{n=0}^{N}\frac{\check\Sigma_{nm}\check\mathcal{I}_{nm}(r_k)}{c_n},
\end{equation}
where $r_k$ denotes the collocation points in the $r$-direction.

Once we obtain $\check\mathcal{I}_{nm}(r_k)$ for $n = 0, 1, \dots, N$,
the calculation of the gravitational potential is trivial.  However,
note that $Q_{m-1/2}(\chi)$ is related to the complete elliptic
integral, which is singular when $r' = r_k$.  In general, this
difficulty arises because of the singularity of
$|\mathbf{x}-\mathbf{x}'|^{-1}$.  In order to avoid the singularity,
we use the ``Chebyshev-roots grid'':
\begin{equation}
  \bar r'_j = \cos\left[\frac{(2j + 1)\pi}{2N}\right],\ \ \ 0 \le j < N
\end{equation}
so that $\hat\mathcal{I}_m(\bar r'_j;r_k)$ is finite for all $m$ and
$j$.  In Figure~\ref{fig:Iknm}, we plot $\hat\mathcal{I}_m(\bar r';r)$
for the first six values of $m$ for $N = 33$, $M = 64$, and $Z(r,z) =
\delta(z)$, which is the case of a very thin disk.

The Chebyshev coefficients $\check\mathcal{I}_{nm}$ are now well
defined as
\begin{equation}
  \hat\mathcal{I}_m(\bar r'_j) = \sum_{n=0}^{N-1}\check\mathcal{I}_{nm}T_n(\bar r'_j)
  = \sum_{n=0}^{N-1}\check\mathcal{I}_{nm}\cos\left[\frac{\pi n(2j+1)}{2N}\right].
\end{equation}
Note that $\check\mathcal{I}_{nm}$ cannot be computed by a standard
discrete cosine transform.  It is related to the discrete Fourier
transform with different parity properties.  This non-standard cosine
transform is known as type-II discrete cosine transform~\citep[see,
e.g.][p.41]{Frigo2003}.  The computational order is still
$\mathcal{O}(N\log_2N)$.  The only potential inconsistency about this
method is that there are only $N$ collocation points, instead of
$N+1$.  Nevertheless, because the Chebyshev coefficients converge
exponentially for smooth functions, $\Sigma_{Nm}$ is assumed to be
very small.  The artificial vanishing of $\check\mathcal{I}_{Nm}$ is,
therefore, negligible in the final solution.

In our implementation, we pre-compute the operator
$\check\mathcal{I}_{km}(r_k)$ to speed up the algorithm.  The steps to
setup the gravity integrator can then be summarized by the following:
\begin{enumerate}
  \item We first take the discrete Fourier transform of
        $\Sigma(r_k,\phi_j)$ along the $\phi$-direction and obtain
        $\hat\Sigma_m(r_l)$, which is of order $\mathcal{O}(NM\log_2
        M)$.
  \item We then calculate the matrix product $\hat\psi_m(r_k) =
        \sum_{l=0}^{N}\hat\Sigma_m(r_l)
        \check\mathcal{I}_{ml}(r_k)/c_n$ for each $m$, which is of
        order $\mathcal{O}(N^2M)$.
  \item Finally, we take the inverse Fourier transform of
        $\hat\psi_m(r_k)$ and obtain the restricted potential
        $\psi(r_k,\phi_j)$, which is of order $\mathcal{O}(NM\log_2
        M)$.
\end{enumerate}
The overall computational cost is $\mathcal{O}(N^2M + NM\log_2M)$.

\subsection{Coupling to Hydrodynamic Equations}

We evolve the hydrodynamic equations following \citet{Chan2005}, i.e.,
using a low-storage, third-order explicit Runge-Kutta method.  The
only difference is that, at the beginning of every time step, we
update the gravitational potential.


\section{Code Verification}\label{sec:tests}

In \citet{Chan2005}, we verified the hydrodynamic part of our
algorithm.  In this second paper, we will first carry out a few
time-independent tests for both the Poisson solver and the direct
gravity integrator.  We use for each test an analytic
density-potential pair and show that the numerical solutions agree
with the analytic expressions.  We also perform some time-dependent
tests, which demonstrate that the gravity solver couples to the
hydrodynamic equation correctly.

\subsection{Time Independent Tests of Poisson Solver}
\label{sec:poisson}

\begin{figure*}
  \plotone{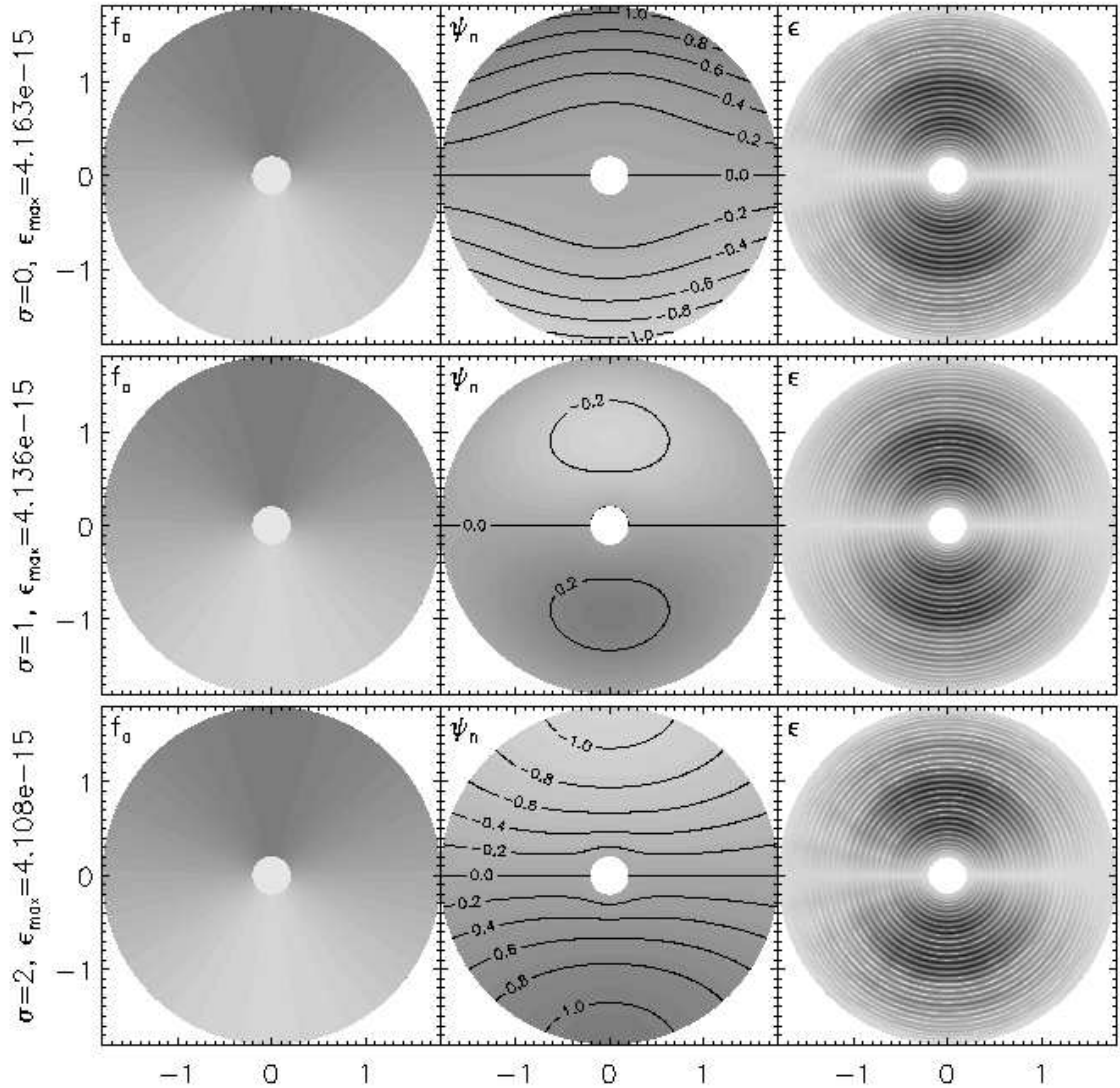}
  \caption{Grayscale plots of the analytical function
           $f_\mathrm{ana}$, the numerical potential
           $\psi_\mathrm{n}$, and the difference $\epsilon$ between
           the numerical solution and the analytical solution for the
           test problem discussed in \S\ref{sec:poisson}.  In all
           plots, darker shades correspond to larger magnitude.}
   \label{fig:poisson}
\end{figure*}

The simplest way to test the Poisson solver is to compare some
numerical potential $\psi_\mathrm{num}$ obtained by the Poisson solver
to its analytical expression $\psi_\mathrm{ana}$.  The tests in this
subsection are done in the computational domain $\mathcal{D}^{(2)} =
[0.2,1.8]\times[-\pi,\pi)$.  We consider a very simple potential
\begin{equation}
  \psi_\mathrm{ana}(r,\phi) = \frac{1}{3}\left[r^2 - \sigma\left(1.82r - \frac{0.0648}{r}\right)\right]\sin\phi,
\end{equation}
where $\sigma$ is some arbitrary parameter.  We chose this potential
because its Laplacian takes the form
\begin{equation}
  f_\mathrm{ana}(r,\phi) = \nabla^2\psi_\mathrm{ana} = \sin\phi,
\end{equation}
which is independent of $\sigma$.  Hence, the parameter $\sigma$
forces our solution to satisfy different boundary conditions.

We choose three different boundary conditions to test the Poisson
solver.  The first one is
\begin{equation}
  \left\{\begin{array}{rcl}
    \psi_\mathrm{ana}(r_{\min}, \phi) & = & \frac{1}{3}r_{\min}^2\sin\phi \\
    \psi_\mathrm{ana}(r_{\max}, \phi) & = & \frac{1}{3}r_{\max}^2\sin\phi
  \end{array}\right.\;,
\end{equation}
which corresponds to $\sigma = 0$.  For the second one, we choose the
vanishing boundary condition
\begin{equation}
  \left\{\begin{array}{rcl}
    \psi_\mathrm{ana}(r_{\min}, \phi) & = & 0 \\
    \psi_\mathrm{ana}(r_{\max}, \phi) & = & 0
  \end{array}\right.\;,
\end{equation}
which corresponds to $\sigma = 1$.  Finally, we also choose
\begin{equation}
  \left\{\begin{array}{rcl}
    \psi_\mathrm{ana}(r_{\min}, \phi) & = & \frac{1}{3}(r_{\min}^2 - 3.64r_{\min} + 0.1296/r_{\min})\sin\phi \\
    \psi_\mathrm{ana}(r_{\max}, \phi) & = & \frac{1}{3}(r_{\max}^2 - 3.64r_{\max} + 0.1296/r_{\max})\sin\phi
  \end{array}\right.\;,
\end{equation}
which corresponds to $\sigma = 2$.  In Figure~\ref{fig:poisson}, we
show the contour of $f_\mathrm{ana}$, $\psi_\mathrm{num}$, and the
difference, $\epsilon = |\psi_\mathrm{num} - \psi_\mathrm{ana}|$, for
each choice of $\sigma$.  The contour lines for $\psi_\mathrm{num}$
show clearly how the boundary conditions change the solution.  The
numerical results agree very well with the analytical solution and the
difference is only of order $10^{-15}$, which is the machine accuracy.

\subsection{Time Independent Tests of Gravity Integrator} \label{sec:int}

\begin{figure*}
  \plotone{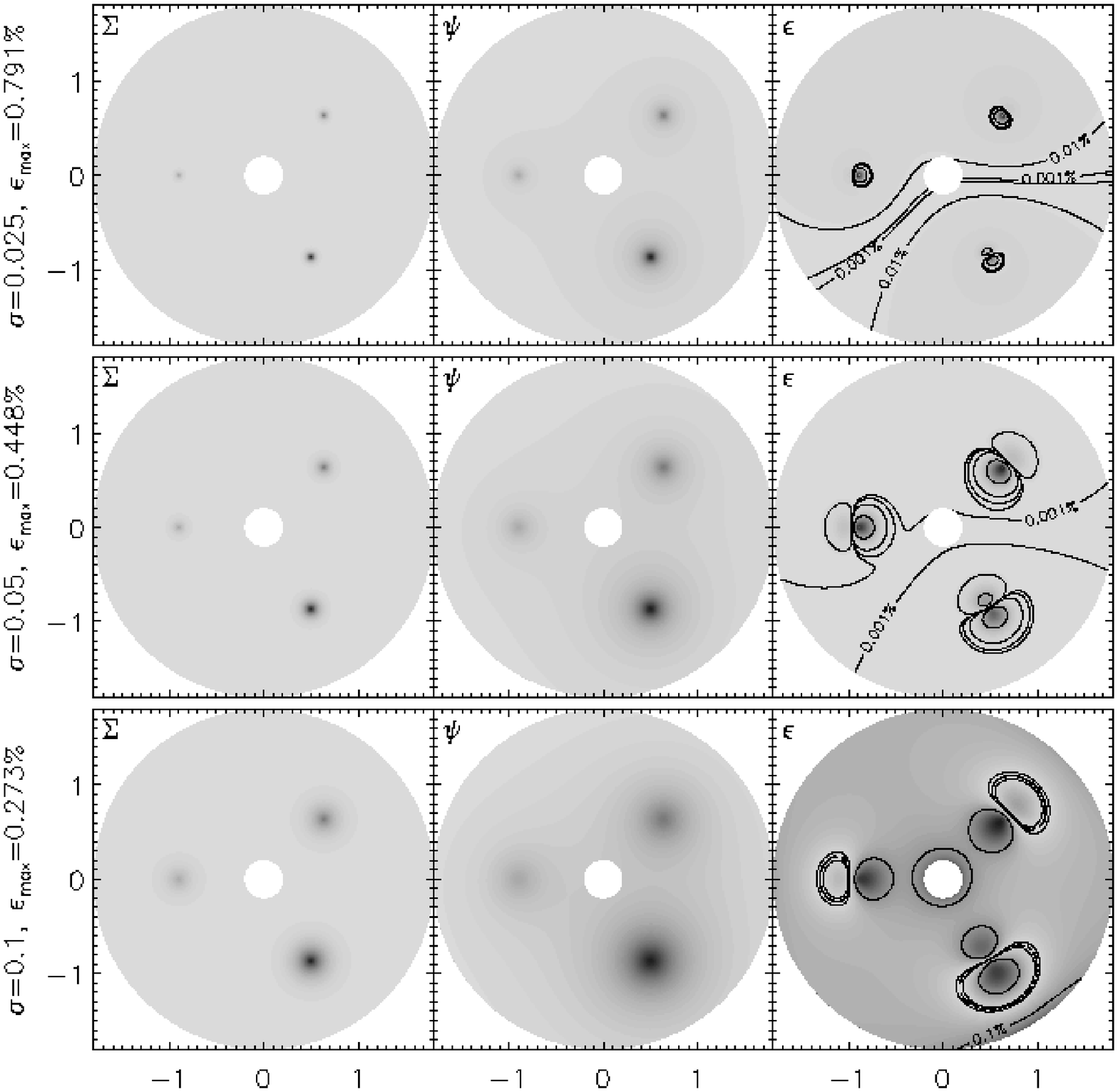}
  \caption{Grayscale plots of the density $\Sigma$, the potential
           $\psi$, and the difference $\epsilon$ between the
           analytical and the numerical values of the gravitational
           potential for a selection of infinitesimally thin
           exponential disks (\S\ref{sec:int}).  The top panel
           corresponds to $\sigma=0.025$, the middle panel corresponds
           to $\sigma=0.05$, and the bottom panel corresponds to
           $\sigma=0.1$.}\label{fig:exponential_disks}
\end{figure*}

\begin{figure*}
  \plotone{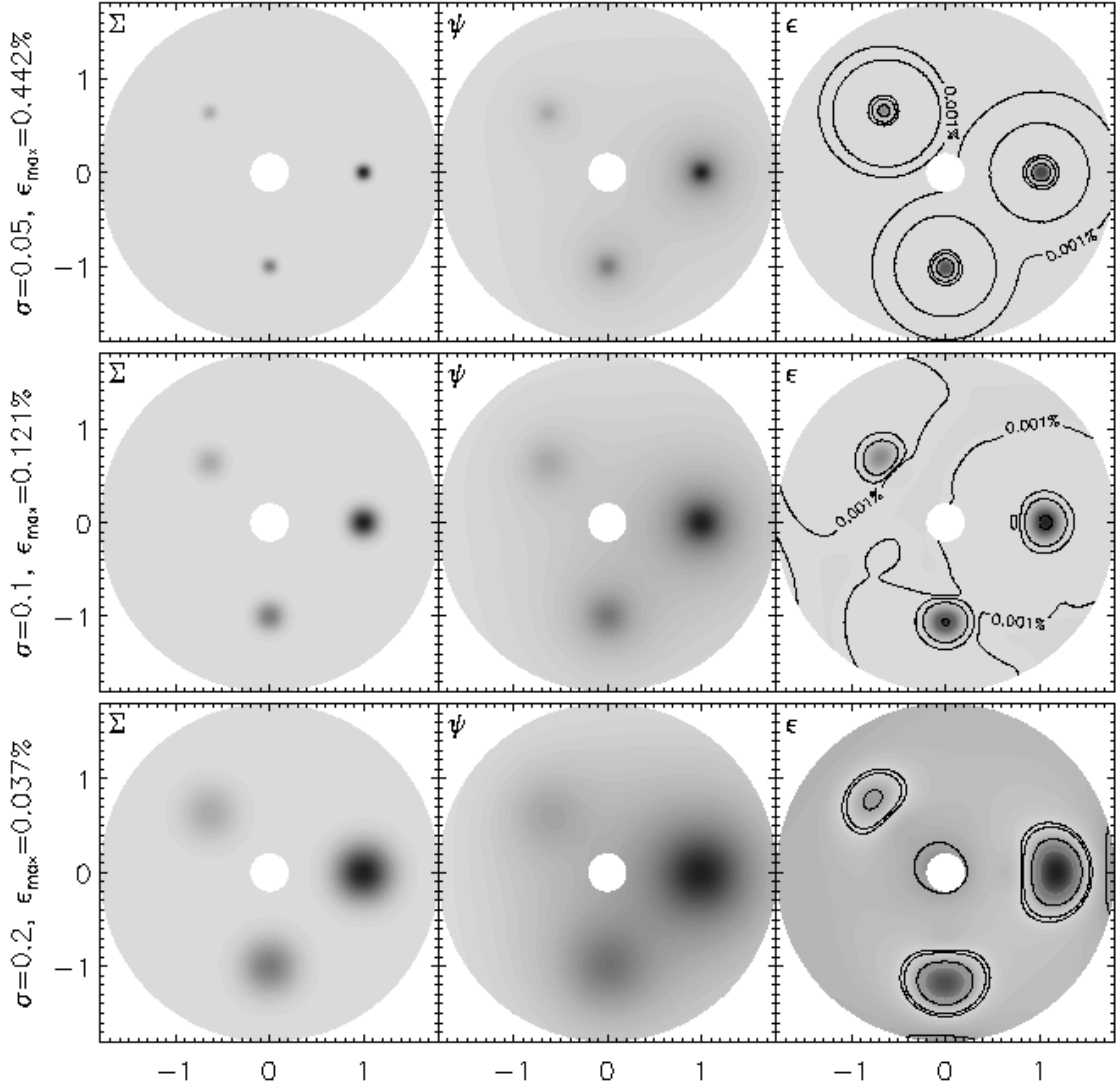}
  \caption{Grayscale plots of the density $\Sigma$, the potential
           $\psi$, and the difference $\epsilon$ between the
           analytical and the numerical value of the gravitational
           potential of a collection of three-dimensional Gaussian
           spheres (\S\ref{sec:int}).  The top panel corresponds to
           $\sigma=0.05$, the middle panel corresponds to
           $\sigma=0.1$, and the bottom panel corresponds to
           $\sigma=0.2$.}\label{fig:gaussian_spheres}
\end{figure*}

To test the gravity integrator, we consider an infinitesimally thin
disk, with an exponentially decaying surface density, i.e.,
\begin{equation}
  \rho(r,\phi,z) = \Sigma(r,\phi)\delta(z) = \Sigma_0 e^{-r/\sigma}\delta(z).
\end{equation}
The corresponding potential on the $z = 0$ plane is given by
\begin{equation}
  \psi(r,\phi) = -\pi G\Sigma_0 r[I_0(y)K_1(y) - I_1(y)K_0(y)]
\end{equation}
where $y \equiv r/2\sigma$ \citep{Binney1987}.  The functions $I_n(y)$
and $K_n(y)$ are the modified Bessel functions of the first and second
kinds, respectively.  The modified Green's function for this problem
is simply
\begin{equation}
  \mathcal{G}(r,\phi,r',\phi') = \frac{-G}{\sqrt{r^2 + r'^2 - 2rr'\cos(\phi-\phi')}}.  \label{eq:green_delta}
\end{equation}

In order to test our gravity integrator for non-axisymmetric problems,
we place three disks in the computational domain with each disk
centered at some $(r_i,\phi_i)$.  Let
\begin{equation}
  R_i \equiv \sqrt{r^2 + r_i^2 - 2rr_i\cos(\phi-\phi_i)} \label{eq:define_R}
\end{equation}
be the distance between $(r_i,\phi_i)$ and $(r,\phi)$.  Normalizing
the total mass for each disk to unity, we have
\begin{equation}
  \Sigma_{(r_i,\phi_i)}(r,\phi) = \frac{1}{2\pi\sigma^2}\exp\left(-\frac{R_i}{\sigma}\right).
\end{equation}
The corresponding potential is
\begin{equation}
  \psi_{(r_i,\phi_i)}(r,\phi) = -\frac{G}{\sigma}y_i[I_0(y_i)K_1(y_i) - I_1(y_i)K_0(y_i)]\;,
\end{equation}
where $y_i \equiv R_i/2\sigma$.

Because the potential is singular, we put the center of each disk off
grid.  Specifically, we choose the coordinates $(0.9,\pi/4)$,
$(0.9,\pi)$, and $(1,-\pi/3)$.  We also change the total mass of each
disk, in order to avoid cancellation of errors by symmetry, by
multiplying them by different constants so that
\begin{equation}
  \Sigma_\mathrm{ana}(r,\phi)
  = \Sigma_{(0.9,\pi/4)}(r,\phi) + \frac{1}{2}\Sigma_{(0.9,\pi)}(r,\phi) + 2\Sigma_{(1,-\pi/3)}(r,\phi).
\end{equation}
We repeated this test for different choices of $\sigma$, namely,
$\sigma = 0.025$, $\sigma = 0.05$, and $\sigma = 0.1$.  In
Figure~\ref{fig:exponential_disks}, we show the grayscale plots for
$\Sigma_\mathrm{ana}$, $\psi_\mathrm{num}$, and the fractional error
$\epsilon = |\psi_\mathrm{num}/\psi_\mathrm{ana} - 1|$, for each value
of $\sigma$.  The contour lines in the error plots show that, at the
smooth density region, the error is of order $10^{-5}$.  The maximum
errors appear around the density peak, where the analytic potential is
singular.

Besides being able to handle a $\delta$-function in the $z-$direction,
our gravity integrator is able to solve problems with other
time-independent and $\phi$-independent vertical structures.  For
example, we consider the Gaussian sphere given by
\begin{equation}
  \rho(r,\phi,z) = \frac{M}{(2\pi\sigma^2)^{3/2}}
  \exp\left[-\frac{r^2+z^2}{2\sigma^2}\right]\;,
  \label{eq:3d_gaussian}
\end{equation}
which is normalized so that the total mass is $M$.  Here $\sigma$ is a
parameter that controls the spread of the mass in each sphere.  As
$\sigma\rightarrow0$, the Gaussian sphere approaches a point source.
If we have a collection of such spheres with the same $\sigma$, the
vertical structure can be factored out of the equations.

Using Gauss' law, we find that the potential on the $z=0$ plane is
given by
\begin{equation}
  \psi(r,\phi) = -\frac{1}{r}\mathrm{erf}\left(\frac{r}{\sqrt{2}\sigma}\right)\;,
\end{equation}
where $\mathrm{erf}(x)$ is the error function
\begin{equation}
  \mathrm{erf}(x) = \frac{2}{\sqrt{\pi}}\int_0^x e^{-x'^2}dx'.
\end{equation}
For a collection of Gaussian spheres centered on the $z = 0$ plane
with the same $\sigma$, we can factor out the $z$-dependence, i.e.
\begin{equation}
  \rho(r,\phi,z) = \sum_i\rho_{(r_i,\phi_i)}(r,\phi,z) = Z(r,z)\sum_i\Sigma_{(r_i,\phi_i)}(r,\phi)\;,
\end{equation}
where we have used the same notation as in the previous subsection to
indicate that the center of each sphere is located at $(r_i,\phi_i)$.
The normalized vertical structure is given by
\begin{equation}
  Z(r,z) = \frac{1}{\sqrt{2\pi\sigma^2}}\exp\left(-\frac{z^2}{2\sigma^2}\right).
\end{equation}
The surface density for each Gaussian sphere is, therefore,
\begin{equation}
  \Sigma_{(r_i,\phi_i)}(r,\phi) = \frac{1}{2\pi\sigma^2}\exp\left(-\frac{R_i^2}{2\sigma^2}\right),
\end{equation}
where $R_i$ is defined in equation~(\ref{eq:define_R}).

The modified Green's function for our problem is
\begin{eqnarray}
  \mathcal{G}(r,\phi;r',\phi') & = & \int_{-\infty}^{\infty}
  \frac{-Z(r',z')dz'}{\sqrt{r^2 + r'^2 - 2rr'\cos(\phi-\phi')+z'^2}} \nonumber\\
  & = & -\frac{e^{R^2/4}K_0(R^2/4)}{\sqrt{2\pi\sigma^2}}\;,
 \label{eq:green_3d_gaussian}
\end{eqnarray}
where $K_0$ denotes the modified Bessel function of the second kind
and we have set $G = 1$ for simplicity.  We use the modified Green's
function described in equation~(\ref{eq:green_3d_gaussian}) to obtain
the corresponding $\check{\mathcal{I}}_{nm}(r_k)$.  Then we follow the
algorithm described in \S\ref{sec:direct_integrator} to perform the
calculation for the gravitational field for three Gaussian spheres
located at $(r,\phi) = (1,0)$, $(0.75, 2\pi/3)$, and $(1.25, -2\pi/3)$
with a total mass of $2$, $1/2$, and $1$, respectively, i.e.,
\begin{equation}
  \Sigma_\mathrm{ana}(r,\phi)
  = 2\Sigma_{(1,0)}(r,\phi) + \frac{1}{2}\Sigma_{(0.9,3\pi/4)}(r,\phi) + \Sigma_{(1,-\pi/2)}(r,\phi)
\end{equation}
In Figure~\ref{fig:gaussian_spheres}, we show the numerical solutions
for $\sigma = 0.05$, $\sigma = 0.1$, and $\sigma = 0.2$.  The maximum
numerical error is less than $0.5\%$.

\subsection{Free Fall of a Dust Ring Under Self-Gravity}

\begin{figure}
  \plotone{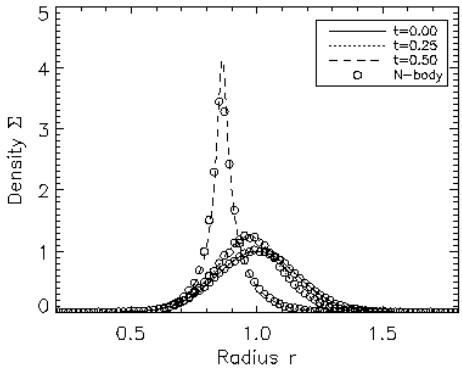}
  \caption{The radial density profile of the numerical solution for a
           free-falling dust ring at $\phi = 0$.  The solid lines
           represent the solution using our spectral algorithm and the
           open circles represent the solution using a simple N-body
           code.} \label{fig:free_fall}
\end{figure}

\begin{figure*}
  \plotone{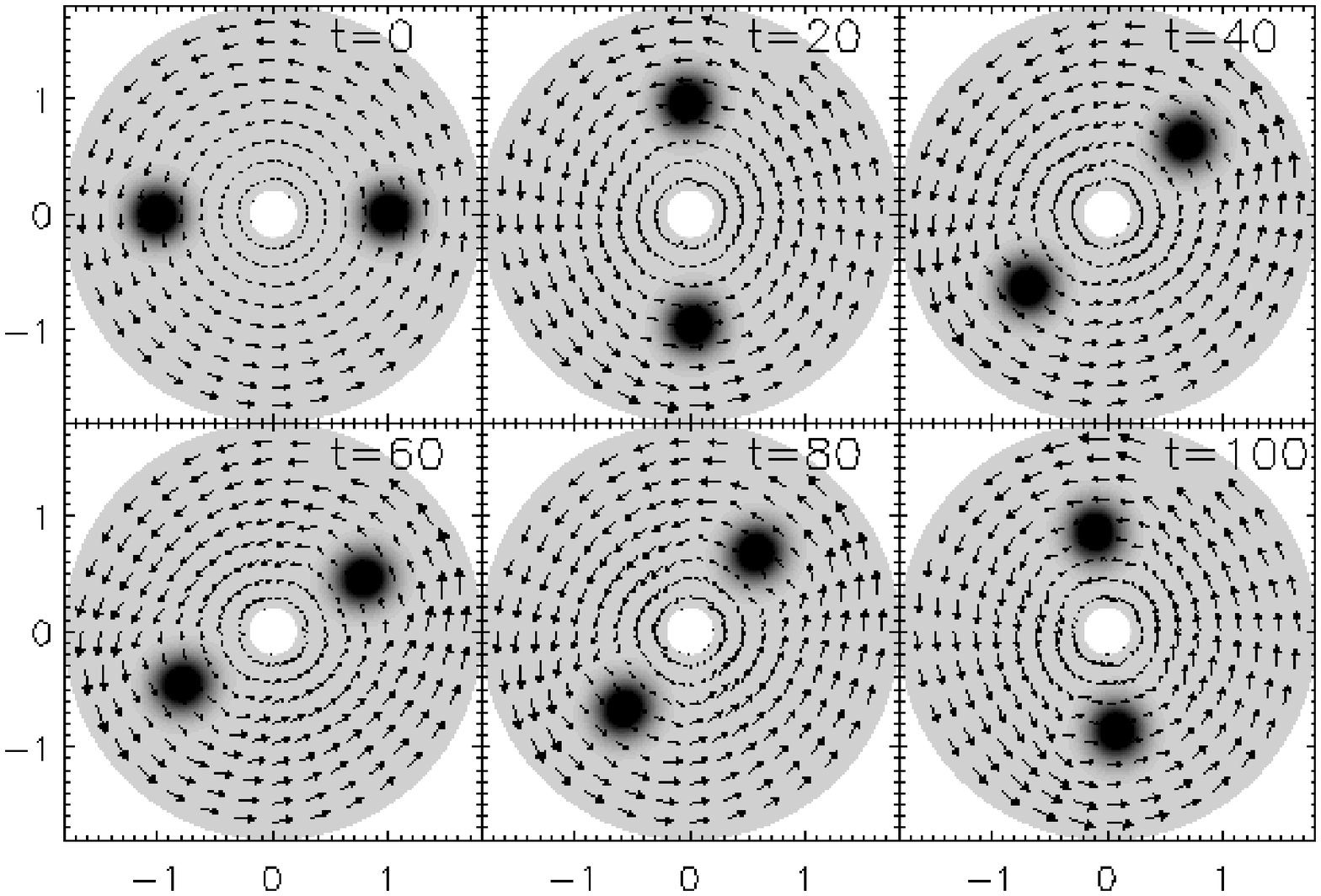}
  \caption{Grayscale plots of the density $\Sigma$ of the two orbiting
           cylinders and vectors plots of the velocity fields at
           different times in the simulation.  In all plots, darker
           colors correspond to larger magnitudes.}
           \label{fig:orbiting_cylinders}
\end{figure*}

In order to perform a dynamic test of our spectral self-gravity
solver, we follow the free-fall of a self-gravitating dust ring.  The
initial condition of this problem is the same as in \S4.1 of
\citet{Chan2005}, i.e.,
\begin{equation}
  \rho(r,\phi,z) = \rho(r,\phi) = \exp[-20(r-1)^2],
\end{equation}
and the initial velocity is zero.  The only difference is that, this
time, we explicitly calculate the self-gravity of the flow instead of
using the gravitational field provided by a central object.  We set
the gravitational constant to $G = 1$ and let the ring free fall.  We
neglect pressure and viscosity and set the resolution to $513\times
64$ in order to resolve the sharp peak in the density profile that
appears at late times.

In order to test our implementation of the spectral method and in the
absence of an analytic solution to the problem we wrote a very simple
N-body algorithm.  We placed 100,000 particles in the computational
domain based on the initial density of the ring.  Then we computed the
gravitational interactions between each particle pair directly and
integrated their trajectories.

In Figure~\ref{fig:free_fall}, we plot the density profiles of the
numerical solution of the free falling dust ring at $\phi = 0$ using
the spectral algorithm and the N-body method.  Different lines
represent the solution from our pseudo-spectral algorithm at different
times and the open circles represent the solution from the particle
method.  The two solutions agree very well.

\subsection{Orbiting Cylinders}

\begin{figure}
  \plotone{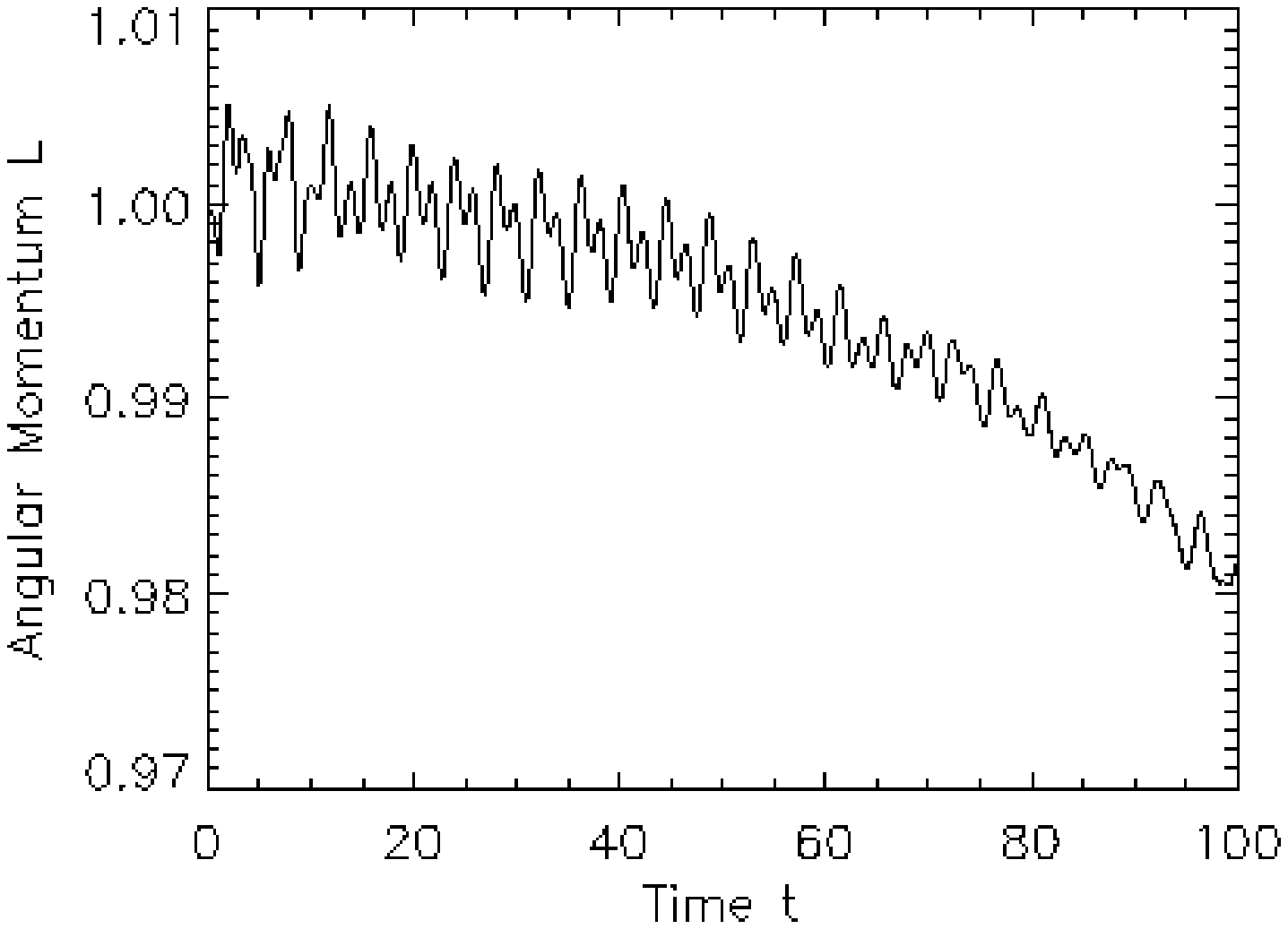}
  \caption{The total angular momentum of the orbiting cylinders
           (normalized initially to unity) as function of time.  The
           angular momentum is conserved to better than 2\% over $t =
           100$, which corresponds to about 16 periods or $\sim
           5\times 10^4$ timesteps.}\label{fig:orbiting_separation}
\end{figure}

We adopt this test from \citet{Fryer2005}, who simulated two spheres
orbiting around each other with smooth particle hydrodynamics.
Because our algorithm is two-dimensional, we assume that all variable
are independent of $z$ and modify the problem to two infinite
cylinders orbiting around each other.

Because the region outside the cylinders has very low density, small
fluctuations of the solution in this low-density region result in a
large error in the pressure force.  In order to resolve the numerical
instability that arises from this effect, we evolve the whole set of
hydrodynamic equations including the energy equation and introduce a
small background term in both the continuity equation and the energy
equation.  This background needs to be small compared to the physical
properties of the cylinders so that it does not affect the orbital
motion significantly, while at the same time it needs to be large
enough so it can screen out the errors at the low density region in
our algorithm.  Then, a strong spectral filter is applied to the
velocity field to reduce numerical instabilities.  The spectral
filters for density and energy, on the other hand, are relatively weak
in order to conserve mass and total energy.

We consider a Gaussian density distribution for each cylinder
\begin{equation}
  \Sigma(r,\phi) = \frac{e^{-r^2/2\sigma^2}}{2\pi\sigma^2}.
\end{equation}
In order to find the initial condition for the energy equation, we
first need to solve for the gravitational potential using the Neumann
boundary condition $\partial\psi/\partial r = 0$ at the origin.  The
second boundary condition is not important here because it only shifts
the potential by a constant, which does not affect the gravitational
field.  Setting the arbitrary constant for the second boundary
condition to zero, we obtain
\begin{equation}
  \psi = G\left[2\ln r - \mathrm{Ei}\left(-\frac{r^2}{2\sigma^2}\right)\right],
\end{equation}
where $\mathrm{Ei}(x)$ is the exponential integral function defined by
\begin{equation}
  \mathrm{Ei}(x) = -\int_x^\infty \frac{e^{-x}}{x'}dx'\;.
\end{equation}
The corresponding gravitational field of this potential is
\begin{equation}
  g(r,\phi) = \frac{2G}{r}\left(e^{-r^2/2\sigma^2}-1\right).
\end{equation}
Therefore, we require the pressure to be
\begin{equation}
  P(r,\phi) \equiv \int \Sigma g\;dr = \frac{G}{2\pi\sigma^2}
  \left[\mathrm{Ei}(-r^2/\sigma^2) - \mathrm{Ei}(-r^2/2\sigma^2)\right]
\end{equation}
in order to balance self-gravity.  Note that we again set the
integration constant to zero in the pressure.  The modified Green's
function for this problem is
\begin{equation}
  \mathcal{G}(r,\phi,r',\phi') =
    G\ln\left[r^2 + r'^2 - 2rr'\cos(\phi-\phi')\right]\;.
\end{equation}

In our simulation we have two Gaussian cylinders centered at
$(r_i,\phi_i)$.  Therefore, we set
\begin{equation}
  \Sigma_{(r_i,\phi_i)}(r,\phi) = \frac{e^{-R_i^2/2\sigma^2}}{2\pi\sigma^2}
\end{equation}
and
\begin{equation}
  P_{(r_i,\phi_i)}(r,\phi) = \frac{G}{2\pi\sigma^2}
  \left[\mathrm{Ei}(-R_i^2/\sigma^2) - \mathrm{Ei}(-R_i^2/2\sigma^2)\right],
\end{equation}
where $R_i$ is defined in equation~(\ref{eq:define_R}).  Because the
pressure is singular when $R_i = 0$, we choose the initial position of
the cylinders $(r_i,\phi_i)$ off grid.  In particular, we also choose
the background density such that the area between the cylinders
contains only $1\%$ of the total mass in the whole system.  Because
the area of our domain is $\pi(1.8^2 - 0.2^2) = 3.2\pi$, we set
\begin{equation}
  \Sigma(r,\phi) = 0.99\left[\Sigma_{(1,10^{-3})}(r,\phi) + \Sigma_{(1,\pi + 10^{-3})}(r,\phi)\right]
  + \frac{0.02}{3.2\pi}
\end{equation}
and
\begin{equation}
  P(r,\phi) = P_{(1,10^{-3})}(r,\phi) + P_{(1,\pi + 10^{-3})}(r,\phi) + \frac{0.02}{3.2\pi}.
\end{equation}
The initial energy is then chosen using the equation $E = 3P/2$ that
is based on an ideal gas law.  Setting $G = 1$, the angular velocity
of the cylinder needed to balance the gravitational force is equal to
unity.  Hence, we set the initial velocities to
\begin{equation}
  v_r = 0, \ \ \ v_\phi = r.
\end{equation}

We use $\sigma = 0.1$ and evolve the simulation to a dimensionless
time $t = 100$, which is equal to about 16 complete orbital rotations.
In Figure~\ref{fig:orbiting_cylinders}, we show the gray-scale plots
of the density as well as the velocity field for different times in
the simulation.  In Figure~\ref{fig:orbiting_separation}, we plot the
angular momentum as a function of time.  Our algorithm is able to
conserve angular momentum to better than 2\% for up to about 16
complete orbiting motion, which corresponds to $\sim 5\times 10^4$
timesteps.  The 2\% dissipation of angular momentum is mostly due to
the strong spectral filtering, which is equivalent to artificial
viscosity, and the deformation of the two cylinders during the
simulation.


\section{The Stability of Self-Gravitating Hydrodynamic Disks}\label{sec:application}

\begin{figure}
  \plotone{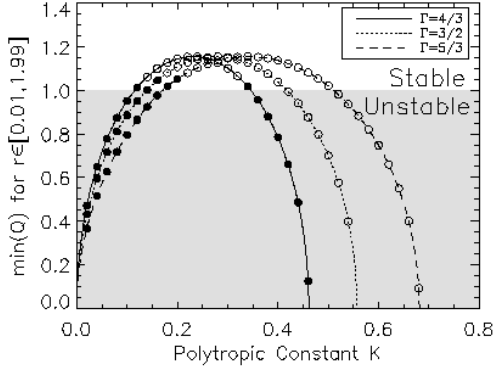}
  \caption{The minimum value of the Toomre parameter $\min(Q)$
           anywhere in the disk, for different values of the
           polytropic constant $K$.  The three different lines
           correspond to different values of the polytropic index
           $\Gamma$.  When $\min(Q) < 1$, i.e.  in the shaded region,
           the linear analysis suggests that the disk is unstable.
           Filled circles denote simulations in which the disk was
           unstable, whereas open circles denote simulations in which
           the disk was stable.}\label{fig:toomre}
\end{figure}

\begin{figure*}
  \plotone{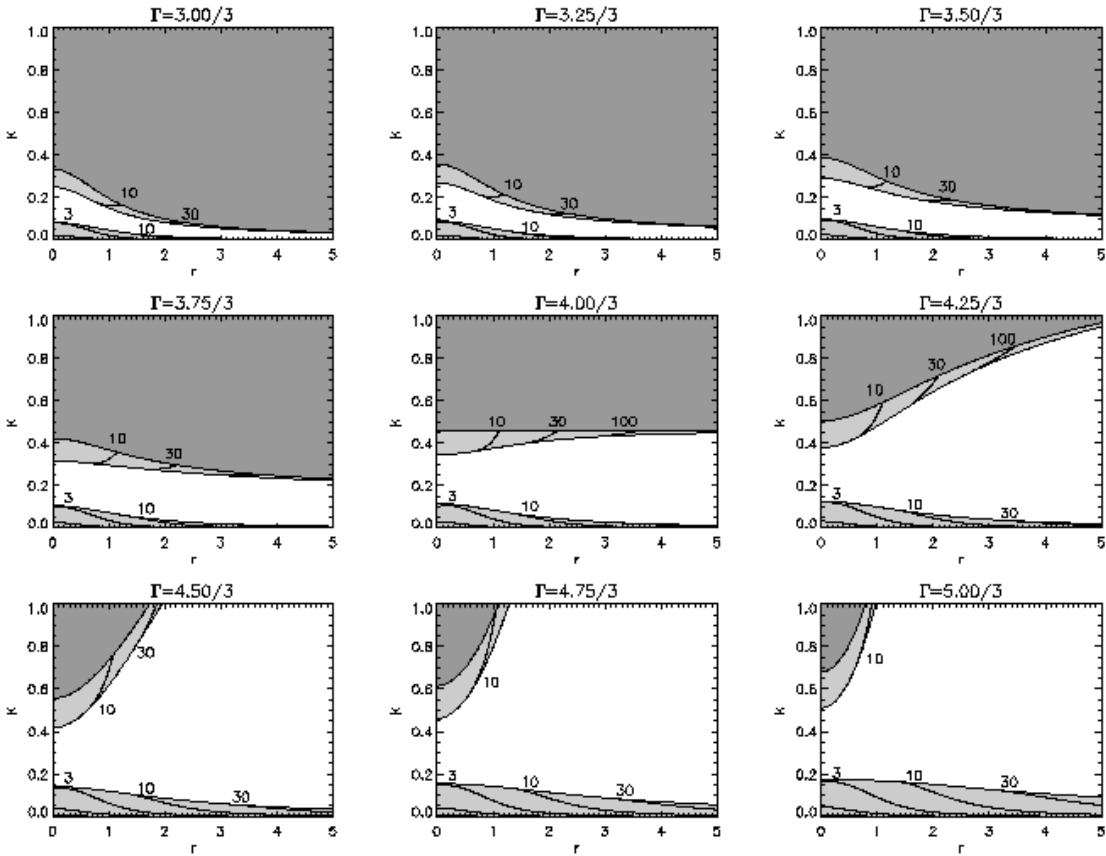}
  \caption{The Toomre parameter $Q^2$ as a function of radius $r$ for
           solutions with different values of the polytropic constant
           $K$.  The dark gray region corresponds to $Q^2 < 0$, for
           which no steady-state rotating solution exists; the light
           gray region corresponds to $0 < Q^2 < 1$ which is Toomre
           unstable; the white region corresponds to $Q^2>1$, which is
           Toomre stable.  The contour lines correspond to the size of
           the minimum unstable wavelength according to the linear
           mode analysis.} \label{fig:q2_kmax}
\end{figure*}

Pseudo-spectral methods effectively evolve in time the hydrodynamic
equations for each mode independently.  For this reason, they are
ideal tools for confirming the results of linear mode analysis and for
extending them to the non-linear regime.  As a case study, we address
here numerically the stability of self-gravitating
infinitesimally-thin disks.

We consider self-gravitating gas disks with a polytropic equation of state
\begin{equation}
  P = K\Sigma^\Gamma, \label{eq:polytropic}
\end{equation}
where $K$ is the polytropic constant and $\Gamma$ is the polytropic
index.  The disks are locally stable to all non-axisymmetric
perturbations \citep{Goldreich1965a,Goldreich1965b,Julian1966}.  For
axisymmetric perturbations, linear-mode analysis of the hydrodynamic
equations results in the dispersion relation \citep[][]{Binney1987}
\begin{equation}
  \omega^2 = \kappa^2 - 2\pi G\Sigma|k| + k^2c_\mathrm{s}^2,
\label{eq:dissipation}
\end{equation}
where we use $c_\mathrm{s}$ to denote the sound speed of the gas,
$\kappa$ to denote the epicyclic frequency in the disk, and $\omega$
and $k$ to denote the mode frequency and wavenumber, respectively.
The disk is locally stable if $\omega^2 > 0$.  The condition for all
modes to be stable is given by Toomre's stability criterion
\citep{Safronov1960, Toomre1964},
\begin{equation}
  Q \equiv \frac{c_\mathrm{s}\kappa}{\pi G\Sigma} > 1\;.  \label{eq:toomre}
\end{equation}
Although criterion~(\ref{eq:toomre}) is only a local condition, it is
also sufficient for global stability of axisymmetric modes with $kr
\gg 1$.  At the critical value $Q = 1$, the disk can be globally
unstable to non-axisymmetric modes.  As $Q$ increases, the disk
becomes globally stable to all non-axisymmetric modes.  However, for
very large values of $Q$, pressure dominates the self-gravity effects
and the disk becomes globally unstable to non-axisymmetric
perturbations again.  Our goal in this section is to study numerically
the stability of self-gravitating disks in situations where the
unstable wavelengths are larger than the characteristic length scales
in the disks.

We simulate the time evolution of disks that obey Plummer's density
model~\citep[see][chapter~2]{Binney1987}
\begin{equation}
  \Sigma(r) = \frac{M_\mathrm{d}\sigma}{2\pi(r^2 + \sigma^2)^{3/2}}\;,
\end{equation}
where $M_\mathrm{d}$ stands for the initial disk mass and $\sigma$ is
a parameter that controls the central concentration of the disk
density.  The corresponding gravitational potential (at the $z=0$
plane) is given by
\begin{equation}
  \psi(r) = - \frac{GM_\mathrm{d}}{\sqrt{r^2 + \sigma^2}}.
\end{equation}
We set the initial radial velocities to zero, i.e., $v_r = 0$, and
choose the azimuthal velocities, $v_\phi$, so as to balance the
pressure and self-gravity of the flow, i.e.,
\begin{equation}
  \frac{v_\phi^2}{r} = \frac{2\pi G\Sigma r}{\sigma} - \frac{3K\Gamma\Sigma^{\Gamma-1}r}{r^2+\sigma^2}.
\end{equation}
Although Plummer's model assumes that the density is a
$\delta$-function along the $z$-direction, its corresponding scale
height $H = rc_\mathrm{s}/v_\phi$ is given by
\begin{equation}
  \frac{H}{r} = \frac{c_\mathrm{s}}{v_\phi} = \left(\frac{2\pi Gr^2}{K\Gamma\Sigma^{\Gamma-2}\sigma}
  - \frac{3r^2}{r^2 + \sigma^2}\right)^{-1/2}.
\end{equation}
We choose $129\times64$ grid points and perform the simulations in the
domain $[0.005,5]\times[-\pi,\pi)$ with $\sigma = 1$.  We set the
total disk mass to $M_\mathrm{d} = 1$.

Figure~\ref{fig:toomre} a parameter study of the dependence of the
stability of the disk on the polytropic index $\Gamma$ and the
constant $K$.  We plot the minimum value of the Toomre parameter
$\min(Q)$ anywhere in the disk against $K$.  The three different lines
correspond to different values of the polytropic index $\Gamma$.
According to linear analysis, when $\min(Q) < 1$, there is some region
in the disk where the disk becomes unstable.  We present the results
of our simulation on the same plot, with filled circles denoting the
solutions that become unstable and with open circles the solutions
that remained stable during the course of the simulations.

For small values of the polytropic index ($\Gamma\le 4/3$) or for
small values of the polytropic constant ($K<0.2$), our numerical
simulations agree with the predictions of the linear mode analysis.
The small disagreements near the $Q=1$ separatrix come from two facts:
that, (i) we have a finite domain of integration on which we have to
impose boundary conditions, and (ii) we cannot calculate the evolution
of the instabilities and the perturbations in the gravitational field
from matter outside the boundaries.  Regarding point (i), we have used
the absorbing boundary condition discussed in the first paper in the
series \citep[see][]{Chan2005}, which reduce the reflection
significantly.  However, a very small but finite amount of the energy
of the wave is still trapped in the domain, which causes the
instability.  As a result, when the gravitational potential deviates
from Plummer's model because of the finite size of the domain, the
disk close to the outer boundaries becomes unstable.

A striking difference, however, appears when the polytropic index and
the polytropic constant become larger: our numerical solutions are
always stable, contrary to the predictions of the linear mode
analysis.  The reason for this disagreement between the linear mode
analysis and our numerical simulations becomes evident in
Figure~\ref{fig:q2_kmax}, where we plot Toomre's parameter $Q^2$ as
function of radius $r$ for different values of the polytropic constant
$K$.  When $Q^2 < 0$, there is no initial value of the azimuthal
velocity $v_\phi$ that satisfies the steady-state equation for radial
force balance; we shade this region using dark gray.  For $0 < Q^2 <
1$, the disk is unstable according to Toomre's criterion; we shade
this region using light gray.  Finally, when $Q^2 > 1$ the disk is
linearly stable; we leave this region white.  We also plot in
Figure~\ref{fig:q2_kmax} contours that correspond to the minimum
unstable wavelength, in the unstable region, computed by
\begin{equation}
  \frac{2\pi}{\lambda_{\min}} = k_{\max} = \frac{\pi\Sigma}{c_\mathrm{s}}\left(1 + \sqrt{1 - Q^2}\right)
\end{equation}

Figure~\ref{fig:q2_kmax} shows that when $\Gamma > 4/3$ and for large
values of the polytropic constant $K$, the minimum unstable wavelength
is larger than the extent of the region in which the disk is unstable.
For example, for $\Gamma = 4.5/3$ and for $K \in [0.42, 0.56]$, the
unstable region only extends to $r \in [0,1.5]$.  However, in that
same region, the minimum unstable wavelength is $\lambda_{\min}
\approx 3$, in the appropriate units.  Because the unstable
wavelengths are larger than the characteristic length scale in the
system (which can be defined here as the extent of the region in which
$0<Q<1$), the approximations involved in the linear mode analysis are
not justified.  Indeed, the numerical simulations show that the
unstable modes cannot grow and the disk is stable.


\section{Conclusions}

Problems that involve self-gravity are usually time-consuming tasks in
computational physics.  In standard finite difference methods,
Poisson's equation is solved as the steady state of a diffusion
equation (using relaxation and over-relaxation methods) and has to be
recalculated together with the hydrodynamic equations at every
timestep.  Although hybrid algorithms have been developed which use
high-order methods to solve Poisson's equation, finite difference
schemes are still used for evolving the hydrodynamic equations.  The
resulting inconsistency in the order of differencing the hydrodynamic
equations and Poisson's equations either significantly reduces the
accuracy of the Poisson solver, or requires the extra complication of
interpolating the gravitational filed at each grid point.  More
importantly, existing two-dimensional hybrid algorithms can only
address a limited number of self-gravitating problems because the
solutions to the two-dimensional and three-dimensional Poisson's
equation are fundamentally different (see \S\ref{sec:introduction} and
\S\ref{sec:equations_assumptions}).

In this paper, we presented two different approaches to using
pseudo-spectral methods to solve self-gravity problems.  For the first
approach, we described the implementation of a standard
pseudo-spectral Poisson solver that solves the two-dimensional
Poisson's equations to machine accuracy.  Instead of solving a
diffusion equation, spectral methods allowed us to invert the Poisson
operator in spectral space, making the algorithm fast and accurate.
For the second approach, we investigated a fast gravity integrator for
disks-like flows with known, time-independent vertical structures.
This algorithm allowed us to study the evolution of flows with finite,
but not only infinitesimal, thickness.  This improvement allows
two-dimensional algorithms to solve a whole new class of problems.  We
demonstrated here the ability of our algorithm to compute properly and
efficiently the gravitational potential of flattened flows using
different test problems.  Even for the high resolution that
corresponds to $129\times256$ collocation points, our Poisson solvers
and integrator use less than 10\% of the total computational time.

We also explored how to extend Toomre's stability criterion to
self-gravitating disks for which the characteristic properties of the
flows change over a length scale that is shorter than the minimum
unstable wavelength.  Based on our simulations, we find that for
Plummer's density model, if the disk is hot and has a polytropic index
$\Gamma > 4/3$, all oscillatory modes in the disk are stable, contrary
to the predictions of the analytic calculation.


\acknowledgements

We thank an anonymous referee for constructive comments that improved
the clarity of our paper.  C.-K.\,C. and D.\,P.\ also acknowledge
support from the NASA ATP grant NAG-513374.


\end{document}